# Quantum Control of Expectation Values of Observables

**Fariel Shafee**
Department of Physics

Princeton University
Princeton, NJ 08540

**Abstract:**

We study the changes if any of the expectation value of a general observable in a quantum system , the difficulties associated with the detection of these changes, and the possible methods for correcting the system through unitary control to maintain a constant average expectation value of the observable.

**Introduction**

Quantum control by laser pulses is now an experimental reality [1] , even though it is not still at a stage where the advantages of coherent methods can be established in a technical or economical sense over conventional noncoherent control. On the other hand, it is only by extending the results obtained so far that the possibilities of application can be explored. The other area of coherent quantum control is in the field of quantum computing. But despite the beauty of the search [2]  or of the  factoring algorithms [3] by a computer built with quantum gates, the field is in a more abstract state of formulation.

Rabitz, Ramakrishna, Schirmer and others [4] have shown in a number of recent papers how sequences of short laser pulses can be used to control the state of a molecule with a chosen Hamiltonian. Unlike dissipative control by similar electromagnetic waves, in these works the control is through a unitary operator representing quantum evolution , and hence reversible. That also makes the formalism far more subtle and often difficult to handle.

And yet the difference between full quantum theoretical and classical methods may sometimes become a bit blurred. The process of measurement is intrinsically involved with control, but there is yet no satisfactory theory of measurement in quantum physics. If decoherence becomes absolutely essential at a stage in any control process, the creation of a unitary procedure in the rest of the process may have to be taken with  a grain of salt. Methods of molecular identification such as spectroscopy have both quantum and classical components.

Nevertheless, if the boundary between the classical and the quantal is moveable [6], the right direction would seem to be to increase the realm of the quantum, because the classical is only the average expectation value of the quantum ensembles, and loses information that might otherwise be put to good use. That is important in control at the molecular level.

In this preliminary paper we try to understand first the implication of the origin of variations in the expectation value of an arbitrary observable, firstly at constant time with the variations present in the intrinsic or environmental  parameters of the system Hamiltonians of different molecules.  Then we look into time variations of expectation values, again arising from intrinsic or environmental sources. Finally we investigate how quantum control may possibly be of use nullify such changes quickly enough so that as a time and ensemble average a reasonably constant expectation value may be maintained.

## Expectation Values of Observables and a Simple Case

In many situations the most important observable of a system is its energy, which is usually an eigenvalue of the Hamiltonian, with the ground state corresponding to the minimal eigenvalue, which must be bounded below. If the Hamiltonian of the system $H_s$ commutes with a number of operators, they will also remain constant in time, as

$$dO/dt = i\,[H,O] \quad\quad\quad [1]$$

for an operator not explicitly dependent on time.

In case symmetry of the $H_s$ under a group, the generators of the group will commute with the Hamiltonian and different state vectors of an irreducible representation may thus have the same energy i.e. there is energy degeneracy.

$$[L_{+-}, H] = 0 \quad\quad\quad [2]$$

giving

$$H\,[L_{+-}\,|E_n\rangle] = E_n\,[L_{+-}\,|E_n\rangle] \quad\quad\quad [3]$$

But $L_{+-}\,|\psi_o\rangle$ may have a different value of some parameter than $|\psi_o\rangle$.

Thus the Hilbert space may have subspaces with different values of the observable but with the same energy For example, in the Bohr model the angular momentum and its projection are degenerate, as the spherically symmetric Coulomb potential commutes with the angular momentum operator. However, the moving electron creates a directed magnetic field and magnetic couplings break the symmetry and the degeneracy in more refined models. In spectroscopy we can see the differences in the energy levels for the differences caused by the perturbing Hamiltonian components. Absorption or emission of electromagnetic waves are affected by all the terms and even the subtlest isotopic mass differences are amenable to measurement by conventional means. NMR identifies the chemical environment of a nucleon spin

The expectation values of such commuting observables may be important in reality but are not particularly interesting for study. If there is no degeneracy and there is a one-t-one correspondence between the observable and energy, we might as well just concentrate on the energy itself and map that into the observable at the end. For example if we have the Hamiltonian

$$H_s = H_s(a,b,c) \quad\quad\quad [4]$$

and

$$\Theta = f(H_s) \quad\quad\quad [5]$$

f being a regular function with a power series expansion and with no further symmetry generator and hence no degeneracy, then

$$\langle\Theta\rangle = f(\langle H_s\rangle) \quad\quad\quad [6]$$

as can be seen trivially.

In this case observing and controlling $<\Theta>$ is equivalent to observing and controlling the expectation value of energy and vice versa. If the ensemble contains molecules with parametric changes in $H_s$ due to different ionizations, or isotopic differences, then that would be reflected in the $<H_s>$ and hence in our $<\Theta>$. Now we can use the sequences of pulses referred to earlier in a straightforward fashion. If the coupling to the system is through a dipole [system] laser light [control field] interaction

$$H_c = - \mathbf{d} \cdot \mathbf{E} \quad [7]$$

$\mathbf{d}$ being the dipole vector and $\mathbf{E}$ the electric vector of the radiation pulse (time dependent), then by changing the frequency of the $\mathbf{E}$ waves we can effectively interact with different molecular classes at different times and that would be reflected in the varying pattern of $<\Theta>$. Let us suppose we have a significant mass dependence [ heavy water versus ordinary e.g.] then the

$$<H_s> = w_1 <H_{s1}> + w_2 <H_{s2}> \quad [8]$$

with $w_1$ and $w_2$ being the weights of two isotopes will give

$$<\Theta> = f(w_1 <H_{s1}> + w_2 <H_{s2}>) \quad [9]$$

and even if f is not a linear function, usually if we know $<H_{s1}>$ and $<H_{s2}>$ [ simple in the ground state] we can work out $w_1$ and $w_2$, as $w_1 + w_2 = 1$.

Here we are interested in a fixed time snap shot of the system and are not using the control $H_c$. So the measurement of $<\Theta>$ has to be done through some other $H_m$.

Actually when we make a single measurement we get a single eigenvalue through decoherence of the system interacting with the macroscopic measuring equipment. To get the expectation value $<\Theta>$ we would have to perform a large number of measurements on clones and we assume that our ensemble contains the such a large number. It is sufficient to make the measurement on a small part of the total system of molecules and still get a statistically reliable $<\Theta>$. The remaining systems can remain as a coherent sea whose prototype has been measured.

For this simple $\Theta$ we do not need $H_c$ to control $<\Theta>$.

## Expectation Value of Observable Not Commuting with $H_s$

If $\Theta$ is not a function of $H_s$ as in the previous case, and in general if it does not commute with $H_s$, we shall have to consider the complete basis of the $\Theta$ eigenvectors and those of the energy separately.

Let an eigenstate $|\theta_n\rangle$ be given in terms of the energy eigenbasis by

$$|\Theta_n\rangle = \Sigma_i\ c_{ni}\ |E_i\rangle \quad \text{...........................................................................................[10]}$$

Let the system be in an arbitrary state $|\psi\rangle$, in general not an eigenstate of $\Theta$. Let

$$|\psi\rangle = \Sigma_k\ a_k\ |\Theta_k\rangle \quad \text{.....................................................................................[11]}$$

Then
$$\langle\Theta\rangle = \Sigma_k\ |a_k|^2\ \Theta_k \quad \text{.................................................................................[12]}$$

If for any reason we change a pair of the coefficients say $a_m$ and $a_n$, then $\langle\Theta\rangle$ will in general change:

$$\Delta\langle\Theta\rangle = 2\ [\ r_m\ (\Delta\ r_m)\ \Theta_m + r_n\ (\Delta\ r_n)\ \Theta_n\ ]$$

But since we must have

$$|r_m|^2 + |r_n|^2 = \text{constant} \quad \text{.........................................................................[13]}$$

we have
$$r_m\ (\Delta\ r_m) + r_n\ (\Delta\ r_n) = 0 \quad \text{..............................................................................[14]}$$

So
$$\Delta\langle\Theta\rangle = 2\ r_m\ (\Delta\ r_m)\ [\Theta_m - \Theta_n\ ] \quad \text{............................................................[15]}$$

There is a phase ambiguity which we are ignoring here for simplicity.

Again, putting [10] into [11] we get

$$|\psi\rangle = \Sigma_{[k,i]}\ a_k\ c_{ki}\ |E_i\rangle \quad \text{.........................................................................[16]}$$

So
$$\langle\Theta\rangle = \Sigma_k\ |\Sigma_i\ a_k\ c_{ki}|^2\ \Theta_k \quad \text{..............................................................[17]}$$

$$= \Sigma_k\ |a_k|^2\ \Sigma_i\ |c_{ki}|^2\ \Theta_k \quad \text{..................................................................[18]}$$

which agrees with Eq(12) on account of the normalization condition:

$$\Sigma_i\ |c_{ki}|^2 = 1 \quad \text{..................................................................................................[19]}$$

Let us now assume that due to perturbations of the system due to intrinsic noncommutativity of $\Theta$ with $H_s$ or due to environmental interaction $H_e$, the state of the system changes. In classical terms we may think first of a population change from one level to another, but in coherent quantum terms we shall speak of this phenomenon as the changes in the coefficients for a pair of particular energy levels. At this stage we are only considering the simpler case of a pair of levels, though obviously this can be generalized to all eigenstates of energy. Let these two levels be *n* and *m*. As a result of this change we shall in general see a change in the expectation value of θ, as in Eq. 15:

$$\Delta <\Theta> = \Sigma_i\ 2\ a_i\ (\Delta\ a_i)\ \ \Theta_i\ \ (|c_{im}|^2+|c_{in}|^2) \quad\quad [20]$$

With the normalization constraint

$$\Sigma_i\ a_i\ (\Delta\ a_i)(\ |c_{im}|^2+|c_{in}|^2\ ) = 0 \quad\quad [21]$$

and the condition that the other energy states are unaffected becomes:

$$\Sigma_i\ (\Delta\ a_i)\ c_{ik} = 0, \quad\quad \text{(for k not equal to m or n)} \quad\quad [22]$$

Again we are not considering the phase ambiguity problem for simplicity. Even with everything real there is an ambiguity in sign.

The same coherent population change between two levels will cause a change in the expected value of $H_s$ :

$$\Delta <H_s> = 2\ \Sigma_i\ a_i\ (\Delta\ a_i)\ (\ |c_{im}|^2\ E_m + |\ c_{in}|^2\ E_n) \quad\quad [23]$$

## Quantum Control of the Expectation Value

Let us first consider the creation of a state with a given $<\Theta>$. The following Lemma, though almost trivial, is a reminder of the difficulty we face:

***Lemma: It is not possible to have a universal unitary operator which transforms an arbitrary state to a sate with a given $<\Theta>$.***

If there is such an operator U then for a given target $|\psi_f>$ and an arbitrary initial state $|\psi_i>$ we must have:

$$\Sigma_n\ U_{mn}\ a_n^i = a_m^f \quad\quad [24]$$

But if U is a unitary matrix it is invertible, and hence we can get any initial vector $\{a_n^i\}$ by operating linearly on the same target vector $\{a_m^f\}$ by the same matrix, which is impossible.

We did not mention whether our basis was the Θ basis or the energy basis, as it makes no difference to the arguments and the conclusion.

However, if the initial state is the energy ground state, and if the potential is such [ e.g. Morse, or square well] such that the difference between any two energy levels has a unique value, then we can proceed as in the Ramakrishna-Rabitz work [ Appendix B] to construct a unitary matrix from a series of pulses:

$$U = \exp[\, i\, \Gamma\,]\, V_n\, V_{(n-1)} \ldots\ldots\ldots V_1 \quad\quad\quad\quad\quad\quad\quad\quad\quad\quad\quad\quad\quad\quad [25]$$

So that the system sequentially jumps from the ground state and picks up the right $\Sigma_k\, a_i\, c_{ik}$ coefficients after passing through all the steps to the last. The situation is similar to the use of Euler rotations to bring a rigid body to an arbitrary rotation from a standard position, each rotation involves only the creation of a new dimension, and here the coefficient of a new eigenvector.

Indeed the rotation analogy is conceptually very illuminating, though we have actually complex unitary matrices in N dimensions with $N^2-1$ parameters, and not orthogonal real matrices with only $N(N-1)/2$ parameters. If we consider the coefficients $a_k$ as real our transformations are only rotations in the N-dimensional space keeping

$$\Sigma_k\, |a_k|^2 = 1 \quad\quad\quad\quad\quad\quad\quad\quad\quad\quad\quad\quad\quad\quad\quad\quad\quad\quad\quad\quad\quad\quad [26]$$

Hence a series of $N^2 -1$ unitary dipole flips of the form

$$V_{mn} = \{\, \exp[i\, \varphi\ |m\rangle\langle n|\,]\,\} \times \{\, \exp[\, -i\, \varphi\ |n\rangle\langle m|\,]\,\} \times \mathbf{1}(\text{remaining N-2 dimensional space}) \quad [27]$$

can produce the requisite unitary transformation. The $\varphi$ depending on the amount of rotation needed and can be generated by the amplitude and phase of the pulse used. The frequency of the electromagnetic wave in the pulse is of course the resonant frequency

$$\hbar\, \omega_{mn} = E_m - E_n \quad\quad\quad\quad\quad\quad\quad\quad\quad\quad\quad\quad\quad\quad\quad\quad\quad\quad\quad [28]$$

Hence the control problem would have been simple if we could at regular intervals bring all the states to the ground state by a universal pull-down operator and then operate on the ground through the required sequence of pulses to regenerate the state with the given $<\Theta>$. But as we found in the lemma, there is no universal unitary pull-down operator. If we need to regenerate from each corrupted state the target state, we would need to know its full description, which is quantum mechanically impossible.

However, for a system with with occasional slow excursions of $<\Theta>$ from the desired value due to partial population exchanges between two energy levels bringning about changes in the weights of different $\Theta$ eigenstates at a time, the syste of Eq. (20-23) are sufficient to solve linearly for the $(\Delta a_i)$'s, from the observed values of $\Delta E$ and $\Delta \Theta$, and the normalization conditions.

Once we get the $\Delta a_i$ we can compensate for them using the pulses as stated earlier, hence regaining the right $<\Theta>$ on average. In general these deviations will be due to emsissions of radiation or relaxation of the higher energy state for transfer to a lower energy state. So we may use a laser pulse with frequency $\omega_{mn}$ and duration sufficient for the right amount of population change from the lower level to the higher one.

Incompatible measurements can be meaningful only in a probabilistic sense [7].

## Conclusions

We have seen that the quantum control of the expectation value of an arbitrary observable is a highly nontrivial problem, with probably no exact solution. Possibly drifts are unavoidable and can only be corrected periodically by laser pulses in an average sense. We intend to numerically work out a few examples to se the extent of control actually attainable.

# Acknowledgement

The author would like to thank H. Rabitz for email discussions.